\documentclass[]{spie}  %>>> use for US letter paper
\usepackage[]{graphicx}
\usepackage{amsmath}
\usepackage{amsfonts}
\usepackage[latin1]{inputenc}
\def \e{\varepsilon}
\def \fl{\rightarrow}
\def \P{{\cal P}}
\title{Collective resonant modes of a meta-surface} 

\author{Didier Felbacq\supit{a}, Emmanuel Rousseau\supit{a} and Emmanuel Kling\supit{b}
\skiplinehalf
\supit{a}Universit\'e de Montpellier 2, Laboratoire Charles Coulomb\\ Unit\'e Mixte de Recherche
du Centre National de la Recherche Scientifique 5221\\ 34095 Montpellier, France; \\
\supit{b}SAGEM, 100 avenue de Paris, 91344 Massy Cedex, France
}

\authorinfo{Further author information: (Send correspondence to D. Felbacq)\\D. Felbacq: E-mail: Didier.Felbacq@univ-montp2.fr}
\begin{document} 
  \maketitle 

%%%%%%%%%%%%%%%%%%%%%%%%%%%%%%%%%%%%%%%%%%%%%%%%%%%%%%%%%%%%% 
\begin{abstract}
A periodic layer of resonant scatterers is considered in the dipolar approximation. An asymptotic expression for the field diffracted is given in terms of an impedance operator. It is shown that surface Bloch modes appear as a collective effect due to the resonances of the scatterers. 
\end{abstract}

%>>>> Include a list of keywords after the abstract 


%%%%%%%%%%%%%%%%%%%%%%%%%%%%%%%%%%%%%%%%%%%%%%%%%%%%%%%%%%%%%
\section{Introduction}
\label{}
Some recent works have shown the interest of considering periodic sets of scatterers forming a two-dimensional pattern on a surface in the low frequency regime. Such structures are called ``metasurfaces'' \cite{metsh}.
In this context, we study the field diffracted by a periodic set of linear nano resonators electromagnetically characterized by their scattering matrix $S$.  %We study the homogenization of this structure %\cite{tartar,bou2,quasi,allaire,nguetseng,kozlov,bensoussan,bou1,alex,alu,silveirinha,F15,F16,F18,F18b,F2,F3,F13}.
We are interested in the regime where the dipolar approximation is valid, but it is assumed that the nano resonators have their internal resonances at wavelengths large enough to lie in this regime. Possible realizations of this model comprises dielectric nano wires with a large enough index (and resonances are Mie-like resonances) or nano-wires doped with quantum dots.
We proceed to an asymptotic analysis\cite{bou} that generalizes the homogenization regime\cite{tartar,bou2,alex,F18b,F13}. The field diffracted by the structure is derived and it is shown that it is characterized by an impedance operator. It therefore behaves as a ``metasurface". 
%In a second step, the strong coupling of this metasurface with a resonant scatterer is studied.
%\section{An asymptotic result}
%The scattering matrix of the nanowires reduces to a single coefficient in the low-frequency limit. However, as we are interested in the resonances that can be supported by the wires
%$q_n=\frac{-1}{1+iR_n}$
%$R_n=\left(\frac{Y_n(ka)}{J_n(ka)}\right)\frac{F(k\sqrt{\e}a-\frac{Y'_n(ka)}{kaY_n(ka)}}
%{F(k\sqrt{\e}a)-\frac{J'_n(ka)}{kaJ_n(ka)}} $, where
%$F(k\sqrt{\e}a)=\frac{J'_n(k\sqrt{\e}a)}{k\sqrt{\e}a J_n(k\sqrt{\e}a)}$
%$S_0 \sim -(\frac{1}{2}i)\pi(F(y)+1/2)x^2$
\section{Expression of the field diffracted by the meta surface}
\begin{figure}
   \begin{center}
   \includegraphics[height=8cm]{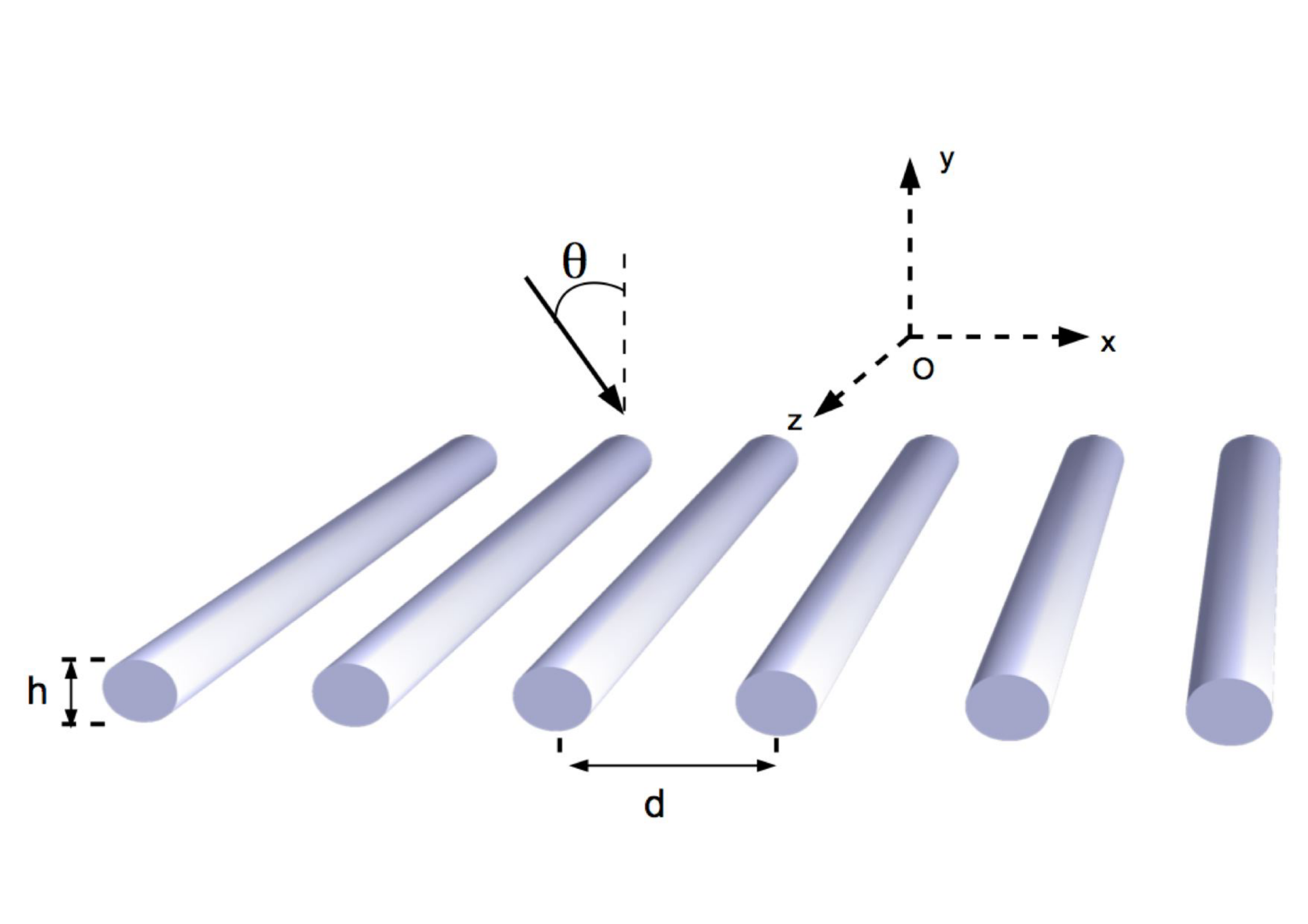}
   \end{center}
   \caption
   { \label{fig1} Sketch of the structure under study.}
\end{figure}

The structure under study (cf. fig. (\ref{fig1})) is made out of an infinite number of resonant scatterers periodically disposed at points $x=p d,y=0$, $d$ is the period and $p\in \mathbf{Z}$. Each scatterer at position $M_p$ is characterized in the frequency domain by a scattering matrix $S(\omega)$ with a non-null component along $Oz$ only. 
The system essentially responds to fields electrically polarized along $Oz$ ($E_{||}$ case of polarization). We therefore consider the case of a $E_{||}$ polarized incident electric field $E^i e^{-i\omega t} e_z$. In order to provide a modal analysis, the incident field is further Fourier transformed along $x$, so that the incident field is a plane wave $U^i(x,y)=e^{i k x\pm \beta y}$, where $k=k_0\sin \theta$ where $\theta$ is an angle of incidence and $k_0=\frac{\omega}{c}$.

For one scatterer alone, the incident field gives rise to a field $U_p^s(M)=\sum_n  s^p_n \varphi_n(M-M_p)$ where $(s^p_n)_n=S(\omega) U^i(M_p)$ and $\varphi_n(r,\theta)=H^{(1)}_n(k_0|r|) e^{i n\theta}$, where $H^{(1)}_n$ is the n$^{th}$ Hankel function of order $1$. For the infinite set of scatterers, this gives a diffracted field that reads as:
\begin{equation}
U^s(M)=\sum_{p,n}  s^p_n \varphi_n(M-M_p).
\end{equation}
Multiple scattering theory allows to write that for $p=0$:
%\begin{equation}
%(b^p_n)_n=S\left( E^i(P_p)+\sum_{m \neq p} T_{pm} (b^m_l)_l \right)
%\end{equation}
$(s^0_n)_n=\left[1-S(\omega) \Sigma(\omega,k)\right]^{-1} S(\omega) U^i(M_0)$, where the lattice sum \cite{lipton} $\Sigma$ is given by: $ \Sigma(\omega,k)=\sum \limits_{\substack{m\neq n}} T_{nm}(s^m_l)_l$, and $T_{nm}$ is a matrix that can be expressed in terms of Hankel functions \cite{josa}.

As the incident field is pseudo-periodic, i.e. $U^i(M_p)=e^{ikd} U^i (M_{p-1})$, then it holds: $\Sigma(\omega,k)=\sum \limits_{\substack{m\neq 0}} e^{ikm}T_{0m}$, ${ k\in [-\frac{\pi}{d},\frac{\pi}{d} ]}$.
%There are many efficient ways of computing this series \cite{lipton}. 
Denoting $\beta_0(\omega,k)=\sqrt{k_0^2-k^2}$, $\beta_n(\omega,k)=\beta_0(\omega,k+n\frac{2\pi}{d})$ and $k_n=k+n\frac{2\pi}{d}$, the scattered field can then be written as a Rayleigh series: 
\begin{equation} \label{rayleigh}
U^{s}(x,y)=\sum_n U^{s}_n e^{i(k_n x+\beta_n(\omega,k) |y|)} \,,
\end{equation}
 where $U^s_n=\frac{2}{d}\sum_l s_l^0 \hat{\phi}_l(k+\frac{2\pi n}{d})$, $\hat{\cdot}$ denotes the Fourier transform along $x$ and Poisson formula was used.
 
% \begin{figure}
%   \begin{center}
%   \includegraphics[height=8cm]{figure1.pdf}
%   \end{center}
%   \caption
%   { \label{fig1} Sketch of the structure under study. It is made of a stack of gratings, each consisting of periodically disposed scatterers.}
%\end{figure} 

The grating can be characterized by the transfer operator \cite{brahim,brahim2,F3}: ${\cal T}(E^s)(x,y)=E^s(x,y+h)$. In the following an asymptotic ($h/\lambda \fl 0$) form of this transfer operator is derived.
%It has a representation as an infinite transfer matrix: $T(E_n^{+})=(E_n^{-})$.
In the limit where the scatterers are very small as compared to the wavelength, the scattering matrix $S(\omega)$ reduces to a scalar matrix $S_0(\omega)$: the scatterers are thus dipoles with a dipole moment along $e_z$. In that case, one obtains:
\begin{equation}\label{s0}
s^0_0(\omega,k)=\left[1-S_0(\omega)\Sigma_0(\omega,k)\right]^{-1}S_0(\omega),
\end{equation} 
where  the series $\Sigma_0$ can be written \cite{bloch,PNFA}:
\begin{eqnarray}\label{sig0}
\Sigma_0(\omega,k)=\sum_{m \neq 0}  e^{ikmd}H_0(k_0|m|d)=-1-\frac{2i}{\pi}\gamma+\frac{2i}{\pi}\ln\left(\frac{4\pi}{k_0 d}\right)+\frac{2}{d\beta_0(\omega,k)}\\
+\frac{2}{d}\sum_{n>0} \left( \frac{1}{\beta_n(\omega,k)}+\frac{1}{\beta_{-n}(\omega,k)}-\frac{d}{i\pi |n|}\right).
\end{eqnarray}

%An asymptotic analysis of this series \cite{bloch} allows to write the following expansion :
%\begin{equation}
%\frac{2}{d}\sum_{n>0} \left( \frac{1}{\beta_n}+\frac{1}{\beta_{-n}}-\frac{d}{i\pi |n|}\right)
%=O[(k_0d)^2]
%\end{equation}
%
Because the only relevant coefficient is $s_0^0$, the grating reduces geometrically to the line $y=0$. 
The scattered field  (\ref{rayleigh}) simplifies to the following form:
%
%\begin{equation} \label{rayleigh2}
%U^{s}(x,y)=\sum_n U^{s}_n e^{i(\alpha_n x+\beta_n |y|)} \,,
%\end{equation}
%where $U^{s}_n=\frac{2 b_0^0}{d \beta_n}$.
%This series splits into a propagative part and an evanescent part:
%{\it Because, we are interested in the low frequency behavior, we expect the evanescent field to be irrelevant.We then obtain the following representation for the scattered field:}
\begin{equation}\label{scatt}
U^s(x,y;\omega,k)=
\frac{2 s^0_0(\omega,k)}{d} \sum_n \frac{1}{\beta_n(\omega,k)} e^{i(k_n x+\beta_n(\omega,k) |y|)}.
\end{equation}
%Let us introduce the Calderòn projector: $F^+=\mathbb{P}^+ F,\, F^-=\mathbb{P}^- F$.
%The boundary conditions can then be written as:
%$$
%F^++F^-=F
%$$
In that expression, there are several propagative waves as well as evanescent waves. The propagative waves correspond to the $\beta_n$'s that are real. We denote by $U=\left\{n\in \mathbb{Z}, \beta_n \in \mathbb{R}^+\right\}$ the set of indices ``n'' corresponding to propagating waves (the diffractive orders of the grating). We are therefore not in the homogenization regime where $k_0 d\ll 1$ and hence there can be several reflected and transmitted orders.

The total electric field reads as:
\begin{equation}
U(x,y)=e^{i(\alpha x-\beta_0 y)}+U^s(x,y) .
\end{equation}
The conservation of energy leads to the following relation (valid above the light cone, i.e. for real $\beta_0$):
$$
|r|^2+\sigma_U\,\Re(r)=0,\, \hbox{ for } \beta_0 \in \mathbb{R}^+,
$$
where: $\sigma_U=(1+\sum \limits_{\substack{n\in U \\ n \neq 0}} \frac{1}{\beta_n})^{-1}$ and $r(\omega,k)=\frac{2 s_0^0(\omega,k)}{d\beta_0(\omega,k)}$. Therefore, above the light cone, $r$ has the following form: $r(\omega,k)=-\sigma_U\frac{1}{1+i\sigma_U \frac{\Im(r)}{|r|^2}}$.

The boundary conditions at $y=0$ are:
$$
[U(x,0^+)-U(x,0^-)]=0,\,  \left[\frac{\partial U}{\partial y}(x,0^+)-\frac{\partial U}{\partial y}(x,0^-)\right]=V(x)\,,
$$
where:
\begin{equation}
U(x)=\sum_n U^{s}_n e^{ik_n x},\, V(x)=2\sum_n i\beta_n U^{s}_n e^{ik_n x}
\end{equation}
%and $r(\alpha)=\frac{2 b_0}{d \beta_0}\sum_n \delta(\alpha-\alpha_n)$. 
These conditions can be rewritten conveniently in the operator form:
\begin{equation}
\left(
\begin{array}{cc}
1 &0 \\
\mathbb{Z} & 1
\end{array}
\right) 
F^+=F^-  \,,
\end{equation}
where:
$F^+=\left( \begin{array}{c} U(x,0^+)  \\ \frac{\partial U}{\partial y}(x,0^+)\end{array}\right)$, $F^-=\left( \begin{array}{c} U(x,0^-)  \\ \frac{\partial U}{\partial y}(x,0^-)\end{array}\right)$ and the impedance operator is defined by: 
\begin{equation}
U(x)=\mathbb{Z}\left[V(x)\right] .
\end{equation}
Conversely, an admittance operator can be defined by: $V(x)=\mathbb{Y}\left[U(x)\right]$.
The operator $\left(
\begin{array}{cc}
1 &0 \\
\mathbb{Z} & 1
\end{array}
\right)$
is the transfer matrix of the meta surface. 
In the homogenization limit of large wavelength $k_0 d \ll 1$, there is only one propagative wave in (\ref{scatt}) which then reads as: $r(\omega,k) e^{ikx+\beta_0y}$. For the propagative part of the field, i.e. if the evanescent field is not taken into account, the transfer operator reduces to a simple $2 \times 2$ matrix:
$$
{\cal T}=\left(
\begin{array}{cc}
1 & 0 \\ 2i\beta_0 r & 1
\end{array}
\right)
$$

%The expression of the reflection coefficient takes a particular form, because of the conservation of energy:
% There exists a function $\chi(k_0,\beta_0)$ such that: \[r(k_0,\beta_0)=\frac{-1}{1+i\chi(k_0,\beta_0)}.\]
% The function $\chi$ is real for real $\beta_0$.
%Energy conservation implies that : $|r|^2+|1+r|^2=1$ and therefore: $\frac{\Re{(r)}}{|r|^2}=-1$, the result follows by defining:
%$\chi=\frac{\Im{(r)}}{|r|^2}$.

%Finally, we determine the Green function $g(\mathbf{r},\mathbf{r}')$. It is the response of the system to a point source $\delta(\mathbf{r}-\mathbf{r}')$. Let $g_0$ denote the Green function in vacuum: $g_0(\mathbf{r},\mathbf{r}')=-\frac{i}{4} H_0^{(1)}(k_0|\mathbf{r}-\mathbf{r}'|)$. From Weyl formula, the plane wave expansion is obtained: $g_0(\mathbf{r},\mathbf{r}')=\frac{1}{4i\pi}\int \frac{1}{\beta_0} e^{i[\alpha (x-x')+\beta |y-y']}d\alpha$. The above derivation leads to:
%\begin{eqnarray}
%g(\mathbf{r},\mathbf{r}';\omega)=g_0(\mathbf{r},\mathbf{r}')+\frac{1}{4i\pi}\int \frac{1}{\beta_0} U^s(x,y;\omega,k)d\alpha, 
%\end{eqnarray}

%The expression under the integral is integrable near $\alpha=\pm k_0$, indeed the quantity $\frac{b_0^0(\omega,k)}{\beta_0}$ is bounded in the vicinity of $\omega=kc$.
%$$
%\Im g(\mathbf{r},\mathbf{r}';\omega)=-1/4 J_0(k_0 |\mathbf{r}-\mathbf{r}'|)-\Im(\frac{1}{i\pi}\int_0^{+\infty} \frac{b_0^0}{d} \sum_n \frac{\cos(\alpha x)}{\beta_n} e^{i\beta_n |y|}d\alpha), 
%$$
%il faut écrire la conservation de l'énergie
%somme des efficacités=1
\section{Scattering properties of the meta surface}
\subsection{Asymptotic  form of the scattering matrix of the nano-wires}
The scattering matrix of the nanowires reduces to a single coefficient $s_0$ in the low-frequency limit. Independently of the nature of the nano wires, the coefficient $s_0$ has a specific form, that originates in energy conservation. Indeed, from the optical theorem \cite{jackson}, the following relation holds: $|s_0|^2+\Re(s_0)=0$. From this, we deduce the following form of $s_0$: \begin{equation}\label{forms} s_0(\omega)=\frac{-1}{1+i\chi(\omega)},\end{equation} where: $\chi(\omega)=\frac{\Im(s_0)}{|s_0|^2}$.

\begin{figure}
   \begin{center}
   \includegraphics[height=8cm]{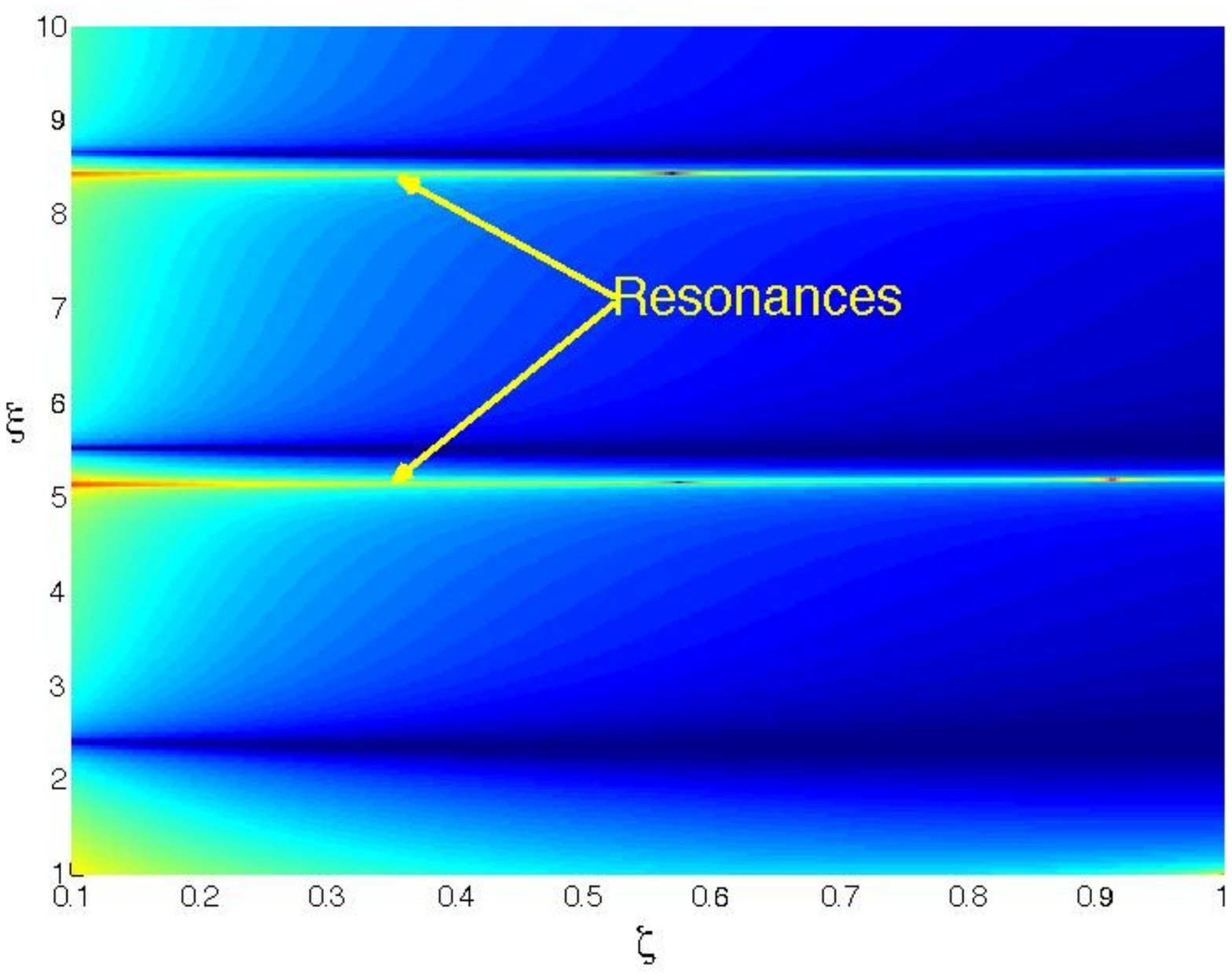}
   \end{center}
   \caption
   { \label{fig2} Modulus of the scattering coefficient $S_0$ as a function of $\zeta$ and $\xi$.}
\end{figure}

We are interested in the resonances that can be supported by the wires, therefore high permittivity values are assumed, so that the condition $k_0\sqrt{\e}a \ll 1$ does not hold. In that situation, the asymptotic study of the $0^{th}$ order scattering matrix is best done by choosing two independent set of variables: $\zeta=k_0a$ and $\xi=k_0\sqrt{\e}a$. For circular nanowires, the expression of $S_0$ is:
$S_0(\zeta,\xi)=\frac{-1}{1+i\chi(\zeta,\xi)}$, where:
$\chi(\zeta,\xi)=\frac{Y_0(\zeta)}{J_0(\zeta)}\frac{F(\xi)-G(\zeta)}{F(\xi)-F(\zeta)}$, and
$F(\xi)=-\frac{J_1(\xi)}{\xi J_0(\xi)},\, G(\zeta)=\frac{Y_1(\zeta)}{\zeta Y_0(\zeta)}$.
The modulus of $S_0(\zeta,\xi)$ is plotted in fig. \ref{fig2}. It can be seen that there are resonances linked to the existence of open cavity modes in the wires. They are distributed along lines that appear in yellow in the figure. When the parameter $\zeta$ is very small, i.e. when the radius of the wire is very small with respect to the wavelength, it can be seen that the resonances are distributed asymptotically along straight lines $\xi=\hbox{\rm cst}$. The resonances are therefore asymptotically invariant under the set of transform $f_{\eta}(a,\e)=(\eta a,\frac{\e}{\eta^2}), \eta >0$. Introducing the dimensionless parameter $\eta$ leads to the asymptotic form: $s_0 \sim \frac{\eta^2\zeta^2}{\eta^2\zeta^2+\frac{2i}{\pi} \frac{\xi J_0(\xi)}{J_1(\xi)}}$. This shows that the resonances of $s_0$ are perturbations to the closed cavity modes given by a Dirichlet condition on the boundary of the nanowire, that is, the zeroes of $J_0$ for a circular cross section. 

\subsection{Bloch modes of the meta-surface}

The Bloch modes that can exist in the structure\cite{F15} are obtained as solutions of the electric field in the absence of an incident field. As the dressed scattering coefficient $s_0^0$ is given by (\ref{s0}), a non-null coefficient implies a pole of $s^0_0$. It is interesting to remark that such a pole cannot be one of $S_0$, because near a pole of $S_0$, one has: $s^0_0 \sim \frac{-1}{\Sigma_0}$ which is bounded. Therefore, modes correspond to zeroes of $1-S_0\Sigma_0$. The dispersion relation is therefore the set of couples $(\omega,k) \in \mathbb{R}^+ \times ]-\frac{\pi}{d},\frac{\pi}{d}]$ such that: \begin{equation} S_0(\omega)\Sigma_0(\omega,k)=1 \,.\end{equation}
 All these modes are necessarily situated below the light cone, as the scattering of a propagating wave is a well-posed problem.
 
\begin{figure}
   \begin{center}
   \includegraphics[height=8cm]{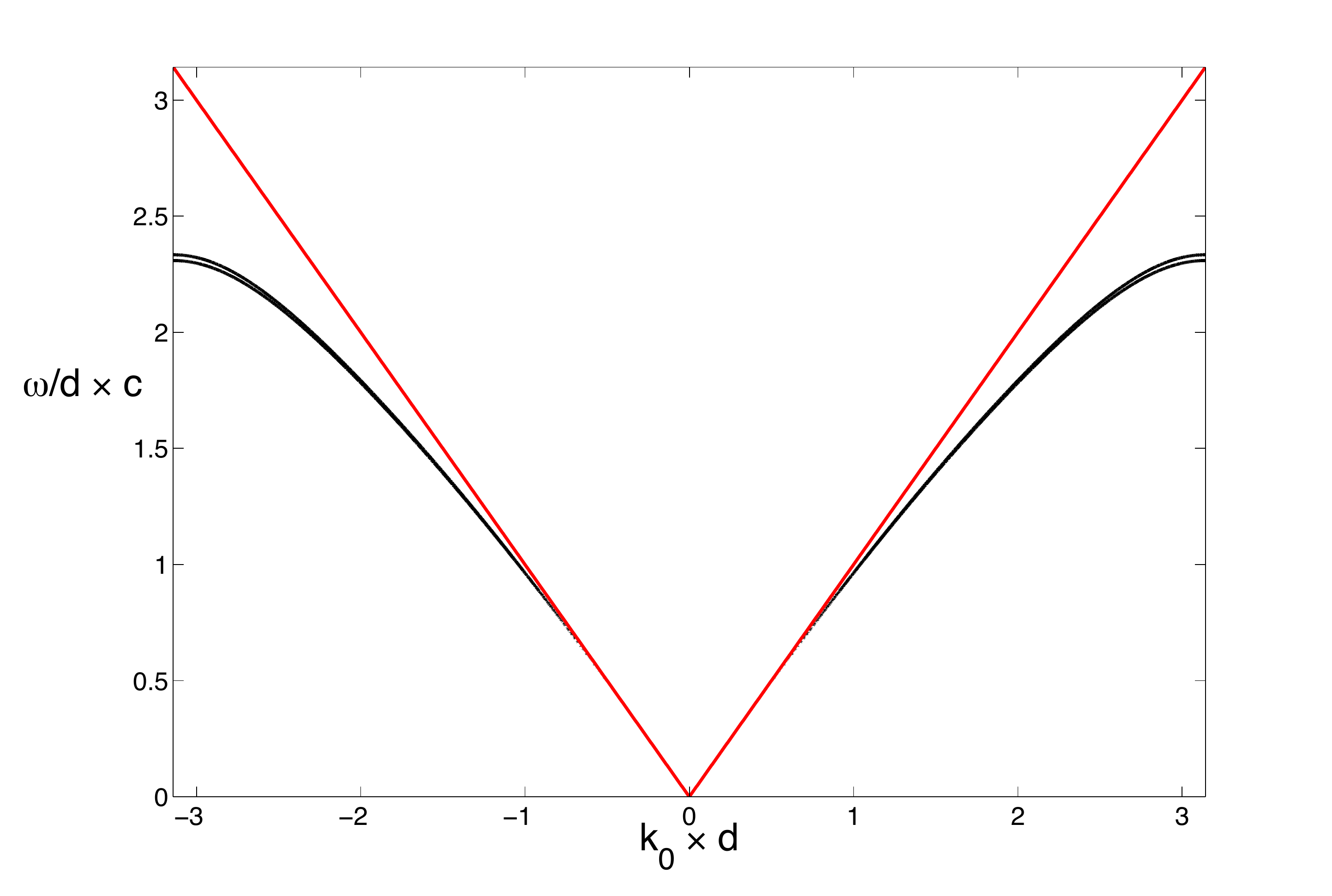}
   \end{center}
   \caption
   { \label{fig3} The dispersion curve for the Bloch modes in the absence of resonances. The light cone is depicted in red.}
\end{figure}

There are two families of Bloch modes. The first one corresponds to the modes that exist in the absence of resonances of the dipoles, i.e. when $S_0(\omega)$ does not have poles. In that case: $S_0(\omega) \sim C k^2_0$ and the expression (\ref{sig0}) for $\Sigma_0$ together with 
an asymptotic analysis of this series \cite{bloch} allows to write the following estimate:
\begin{equation}
\frac{2}{d}\sum_{n>0} \left( \frac{1}{\beta_n}+\frac{1}{\beta_{-n}}-\frac{d}{i\pi |n|}\right)
=O[(k_0d)^2]
\end{equation}
This leads to the following crude approximate form $\Sigma_0 \sim \frac{2}{d\beta_0(\omega,k)}$, from which we derive the following approximate dispersion relation near the $\Gamma$ point:
$$
k_0 \sim \frac{d}{2\sqrt{2}C}\sqrt{1+\sqrt{1 - \frac{16 C^2}{d^2} k^2}}
$$
This family of Bloch modes is illustrated in fig.\ref{fig3}.

\begin{figure}
   \begin{center}
   \includegraphics[height=8cm]{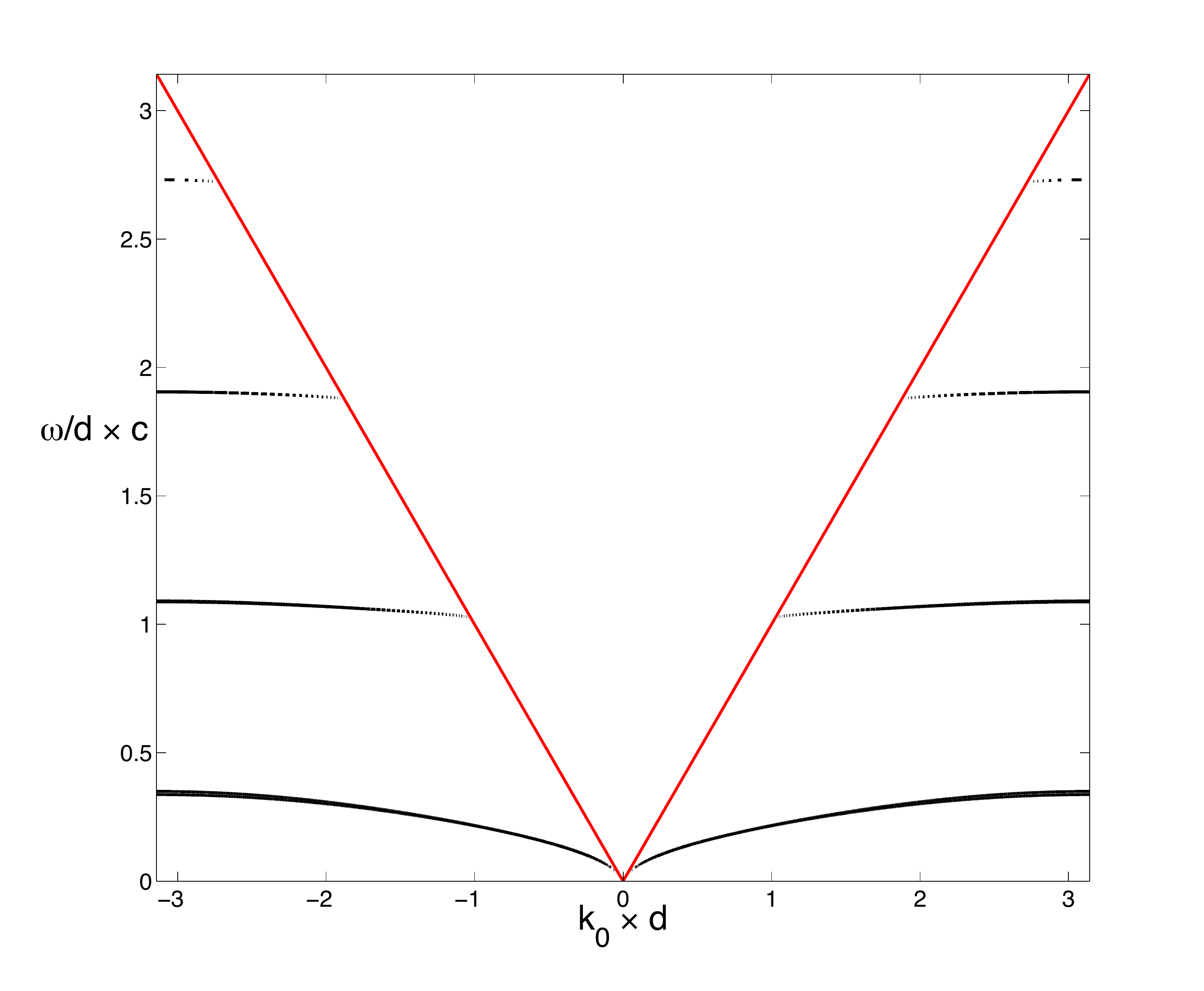}
   \end{center}
   \caption
   { \label{fig4} The dispersion curve of the Bloch modes when several resonances are present. }
\end{figure}

\begin{figure}
   \begin{center}
   \includegraphics[height=8cm]{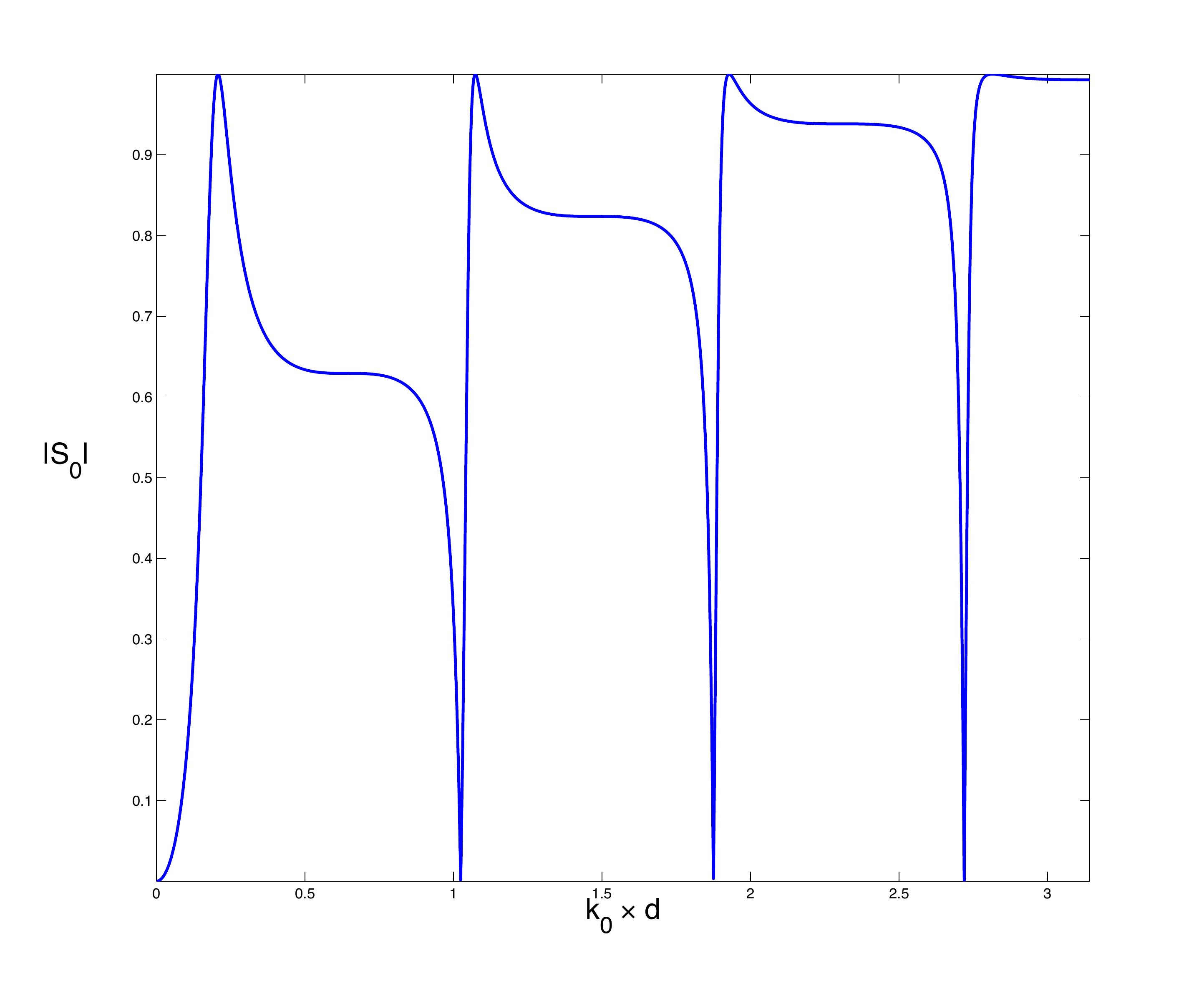}
   \end{center}
   \caption
   { \label{fig5} Modulus of the scattering coefficient $S_0$. }
\end{figure}

The second family corresponds to Bloch modes which arise from the resonances of each dipole, forming bands due to the coupling between them, leading to spatial dispersion\cite{F15}. The particular form (\ref{forms}) of $S_0$, which results, as it has already been said, from energy conservation, has important consequences: if $S_0$ had no zero then $\chi$ would be bounded and by Liouville theorem it would be constant. If $\chi$ had no zero, then by considering $(1-S_0)^{-1}$ we would conclude that $S_0$ is constant. 
Consequently, $\chi(\omega)$ has, locally, the form $\chi(\omega)=\frac{\omega-\omega_z}{\omega-\omega_p}G(\omega)$, where $G$ is regular, and therefore: $S_0(\omega)=-\left(1+i \frac{\omega-\omega_z}{\omega-\omega_p}G(\omega)\right) ^{-1}$. As a rule, poles are therefore associated with zeros.
In that case, near a resonance, one has in the vicinity of a pole $\omega_p$ of $S_0(\omega)$: $S_0(\omega)\sim \kappa \frac{\omega-\omega_z}{\omega-\omega_p}$ and the dispersion relation becomes:
\begin{equation*}
\kappa \times (\omega-\omega_z)\Sigma_0(\omega,k)=\omega-\omega_p.
\end{equation*}
Because $\Sigma_0(\omega,k)$ is singular along $\omega=k_0\, c$, the dispersion curve has to end at the point $(\frac{\omega_z}{c}, \omega_z)$. The collective modes are therefore linked to the zeros of $S_0(\omega)$ rather than the poles, although both are tightly connected as proven above.
This is illustrated in fig. \ref{fig4} where we have plotted the dispersion curves when several resonances are present. In fig.\ref{fig5}, we have plotted the modulus of $S_0(\omega)$. It can be verified that the dispersion curves end on the light cone at the zeros of $S_0(\omega)$.
\subsection{Scattering of an incident plane wave}
\begin{figure}
   \begin{center}
   \includegraphics[height=8cm]{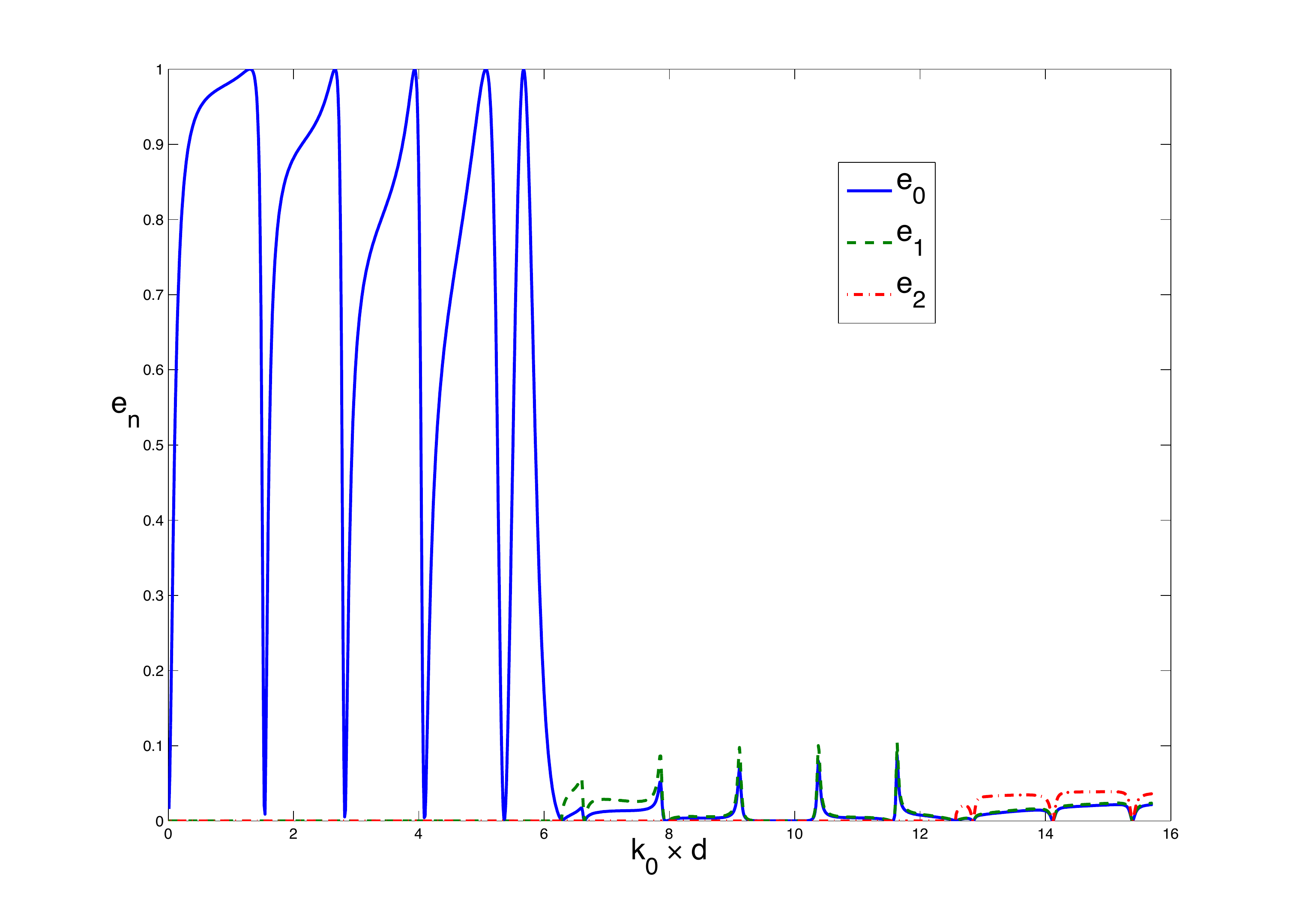}
   \end{center}
   \caption
   { \label{fig6} Energy diffracted in the different refracted orders for resonant scatterers. }
\end{figure}

\begin{figure}
   \begin{center}
   \includegraphics[height=8cm]{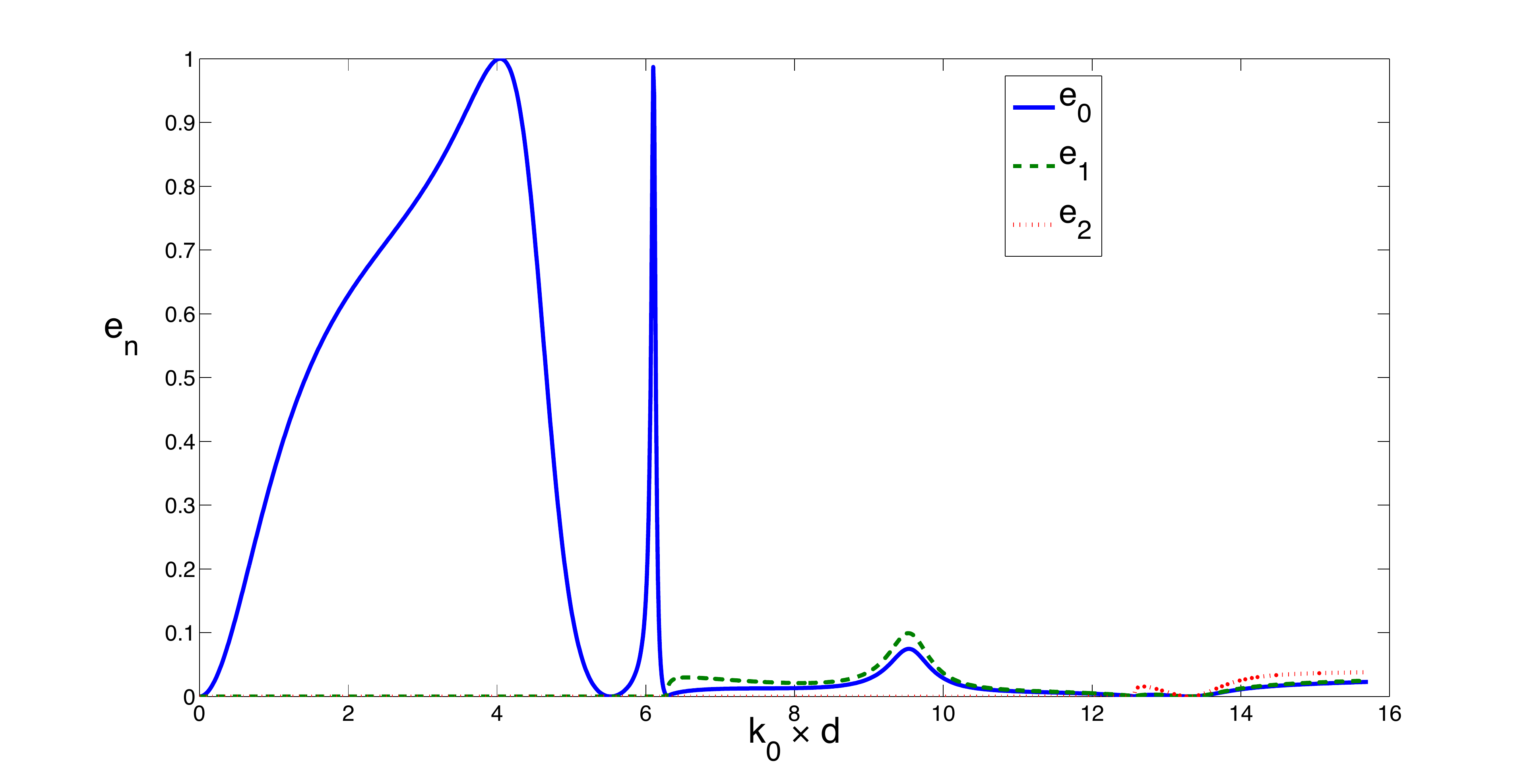}
   \end{center}
   \caption
   { \label{fig7} Energy diffracted in the different refracted orders for non resonant scatterers. }
\end{figure}

As it was demonstrated above, the field diffracted for an incident plane wave is given by eq. (\ref{scatt}). The Poynting vector $\P$ is given, up to an irrelevant factor, by: $\Im{\left(U \nabla \bar{U}\right)}$. Considering a horizontal segment situated at $y=y_0$ and extending over one period, the flux $\Phi$ of $\P$ through this segment normalized by the flux $\Phi^i$ of the Poynting vector of the incident field is given by: $$\frac{\Phi}{\Phi^i}=-1+\frac{4 |s_0^0|^2}{d}\sum_{n\in U} \frac{1}{\beta_n}$$ This shows that the reflected efficiencies (in the meaning of grating theory\cite{brahim,brahim2}) are $e_n=\frac{4 |s_0^0|^2}{d \beta_n}$. In fig. \ref{fig6}, we have plotted the energy in the diffracted orders for an incident plane wave in normal incidence. It can be seen there that, when there is one diffracted order only (i.e. $\lambda \geq 2d$), the overall structure retains the zeros of the scattering coefficient but also that it is perfectly reflective at wavelengths close to that where the modulus of the scattering coefficient $S_0$ is equal to $1$. For smaller wavelength, other diffracted orders appear. In the region where there are $2$ orders, it can be seen that the efficiencies are sharply peaked and that these orders transport about the same energy. The meta surface has therefore the capacity of refracting (or transmitting) energy in two different directions with the same efficiency. Moreover, the wavelength at which this transfer takes place can be tailored by modifying the properties of the nano-wires. In fig. (\ref{fig7}) we have plotted the efficiencies in the situation of a meta surface made of non resonant scatterers. When there is only one reflected order, the phenomena of complete reflection and complete transmission are still present. 

\section{Conclusion}
We have described an approach to the asymptotic description of a one dimensional metasurface consisting in replacing a grating of nano wires by an impedance condition. This led to an explicit form of the transfer matrix and fine results on the Bloch modes, with the evidence of the existence of bands linked to collective resonances propagating along the surface. It would be possible to generalize the study to non-periodic, for instance quasi-periodic, structures \cite{quasi}.
This asymptotic study could also be extended to deal with the new emerging field of quantum metamaterials \cite{newsroom}, as for periodic atomic lattices \cite{mauro1,mauro2}, with the study of the strong coupling regime of an emitter with the Bloch modes along the metasurface \cite{caze}. Moreover,  the impedance matrix formalism presented here could be used to study out-of-equilibrium properties \cite{riccardo}: an interesting direction of research would be the study of the quantum thermalization mecanism for a quantum system close to such a meta surface \cite{mauro3,mauro4,rous1,rous2}. This opens the way to thermal control by means of meta surfaces.
\vskip 1cm
{\bf Acknowledgments}\\
The financial support of the Agence Nationale de la Recherche through grant 060954 OPTRANS is acknowledged. D. Felbacq is a member of the Institut Universitaire de France.

\end{document}